\documentclass[%
pra,
amsfonts,
floatfix,
frontmatterverbose, 
]{revtex4}

\usepackage{subfigure}
\usepackage{graphicx}
\usepackage{dcolumn}
\usepackage{bm}
\usepackage{amssymb}
\usepackage{amsmath}

\begin{document}

\title{Entanglement dynamics between two-level atoms surrounding a microtoroidal cavity and influences of initial states}

\author{Emilio H. S. Sousa}
 \email{ehssousa@ifi.unicamp.br}

\author{J. A. Roversi}%
 \affiliation{%
 Instituto de Fisica "Gleb Wataghin", Universidade Estadual de Campinas, 13083970 Campinas, SP, Brazil
}%

\date{\today}

\begin{abstract}
We theoretically investigate how the initial state influence the entanglement dynamics between two and three two-level atoms with dipole-dipole interaction (DDI) coupled to a whispering-gallery-mode (WGM) microtoroidal cavity. Two different cases, where the two atoms are coupled symmetrically or asymmetrically to the two WGMs through evanescent fields, are discussed in detail. Considering two types of initial states between the atoms and the symmetric regime, we show that for the initial entangled state, the sudden death and birth, as well as the freezing of the entanglement, can be obtained by adjusting both the scattering strength between the modes and the DDI, differently from the initial product state. Moreover, we note that the atomic entanglement generation is more susceptible to the scattering strength variation between the modes than to the DDI. In addition, for the asymmetric regime, the entanglement generation is strongly dependent on the atomic location and the scattering strength. Similar results are obtained for the case of three atoms coupled to a microtoroidal cavity, even in the presence of losses.

\end{abstract}

\pacs{Valid PACS appear here}
\maketitle


\section{Introduction}
\label{intro}

Entanglement, a striking characteristic of quantum mechanics, plays an important role in quantum information processing including quantum cryptography \cite{Ekert}, quantum computation \cite{Gottesman} and quantum teleportation \cite{Bennett}. In the past few years, a lot of effort has been invested into the controlled generation and conservation of entanglement in various quantum systems, such as cavity quantum electrodynamics (CQED) \cite{Pellizzari,Ye,Miller,Boozer}, atomic ensemble \cite{Thompson}, ion traps \cite{Wineland} and superconducting systems \cite{Riste}. In particular, CQED systems promote an efficient route towards quantum control of strong coherent interactions between atoms and photons, offering new opportunities to explore quantum communication tasks. However, the conventional Fabry-Perot cavity presents limitations that makes difficult to build large-scale quantum communication networks due to the leakage caused by this type of cavity. On the other hand, recently, a number of new architectures for CQED systems have been proposed as an alternative for the conventional single-mode CQED (Fabry-Perot) \cite{Vahala,Kimble}. Among these cavities, the microtoroidal cavity \cite{Vernooy,Armani,Spillane} has been attracting considerable attention to the investigation of fundamental physical processes, ranging from the basic study CQED phenomena \cite{Sequin} to nonlinear optics \cite{H.Kimble}. This optical resonator supports whispering-gallery modes (WGMs) which allow ultrahigh quality factors, very small mode volumes and the efficiency enhancement of light-matter interaction \cite{Vahala}. Unlike the conventional Fabry-Perot cavity, the microtoroidal resonator typically includes two counterpropagating WGMs, i. e., clockwise (CW) and counterclockwise (CCW) propagating modes, which couple to each other due to backscattering induced by cavity imperfections or surface roughness \cite{Yi,Jia}. These two WGMs have the same polarization and a degenerate frequency. Near the surface of the resonator, dipole emitters (an atom, a quantum dot or a diamond NV center) are able to interact with the two WGMs through the evanescent fields. With a tapered fiber waveguide coupling the WGMs, the efficiency for coupling quantum fields into and out of the microtoroidal cavity can approach 0.99-0.999 \cite{Painter,Kippenberg}. 

Based on the strong coupling between the microresonators and atoms, some schemes about single photons transistor \cite{Hong}, photon turnstiles \cite{Dayan,Liu}, quantum controlled-phase-flip gates \cite{Xiao,Jin}, optical switching \cite{Shea,Parkins} and photons routers \cite{Aoki,Shomroni} have been theoretically and experimentally reported. Taking into account the single-photon transport properties which are well determined by the fiber-tapered-cavity coupling and, strong atom-field coupling, the electromagnetically induced transparency (EIT) phenomena have been observed extensively due to the detuning between field frequency and atomic transition frequency \cite{Shen}. Another important system, a fiber-coupled microtoroidal cavity which interacts with multiple nanoparticles (which can be seen quantum emitters), has attracted great interest \cite{Yuecheng,Astratov,Zhu,Cordoba}. In Ref. \cite{Chen}, the authors show by means of the transmission spectrum how to extract information of the system via nanoparticle sensing using the WGMs in the large particle-influx regime. In Ref.\cite{Yu} the authors use multiple nitrogen-vacancy centers each embedded inside a single diamond nanocrystal interacting with the WGMs microcavity to explore the effects of the Rayleigh scattering of the nanocrystal in generating entanglement between two nitrogen-vacancy centers.

Despite these accomplishments, to our knowledge, the influence of the initial state under the entanglement dynamics of the composite system containing multiple atoms (or dipole emitters) with DDI and coupled to a microtoroidal cavity has not yet been well studied. In this work, we pay attention to how the initial state influences under the entanglement dynamics between two and three two-level atoms with DDI coupled to a microtoroidal cavity. 
Two special cases were considered: one is that the atoms are coupled symmetrically to the two WGMs, and the other is that the atoms are coupled asymmetrically to the two WGMs, both through evanescent fields. We discuss the time evolution of the atomic entanglement associated with two different initial states, i.e., entangled and product state between the atoms. It is shown that depending on the atomic location and the system parameters, the initial state has a notable contribution on the entanglement dynamics between two and three atoms coupled to a microtoroidal cavity. In the case where the two atoms are initially prepared in an entangled state and symmetric regime, the sudden death and birth, as well as the freezing of the entanglement, can be obtained by adjusting both the scattering strength between the modes and the DDI. The maximal entanglement is more susceptible to variations in scattering strength than the DDI. For the initial product state between the atoms, the entanglement dynamics between the two atoms is dependent on the atomic location and the scattering strength in the asymmetric regime. Taking into account the cavity leakage and atomic spontaneous emission, we find that the decoherence in the entanglement dynamics between two and three atoms can be compensated  by adjusting both the scattering strength and the DDI for the two types of initial states considered.
 
Our paper is organized as follows. First, in Section \ref{sec:model}, we present the model and description of the physical system. 
In Section \ref{sec:result1}, we show and discuss the entanglement dynamics between two two-level atoms for different initial states and parameters system. In Section  \ref{sec:result2}, we introduce the dissipation effects on entanglement dynamics of the system. In Section  \ref{sec:multipartite}, we present the case for tripartite atomic entanglement with dissipation. Finally, we summarize our conclusions in Section \ref{sec:conclusion}. 

\section{Theoretical model}
\label{sec:model}

\begin{figure}[htp]
\centering
\resizebox{0.33\hsize}{!}{\includegraphics*{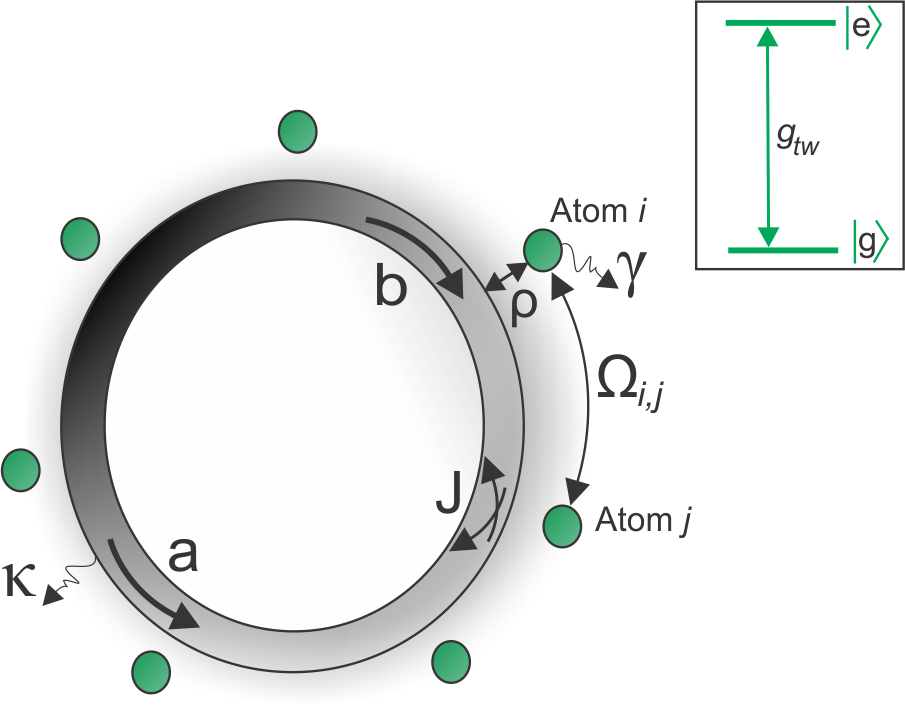}}
\caption{(Color online) Schematic representation of the system. The system consists of a WGM microtoroidal resonator (gray toroid) coupled to multiple identical two-level atoms (green sphere) with DDI represented by $\Omega$. The coupling between the multiple atoms and each WGM is given by $g_{tw}$. The scattering strength between the two WGMs due to the cavity imperfection is $J$. The intrinsic loss of each WGM and each atom is $\kappa$ and $\gamma$, respectively. The inset presents the level configuration for the atoms.}
\label{fig:fig01}
\end{figure}

Schematic description of the composite system which consists of a microtoroidal cavity and multiple two-level atoms with DDI, is shown in Figure \ref{fig:fig01}. The microtoroidal cavity supports a pair of WGMs which are described in terms of the annihilation (creation) operators $\hat{a}$ ($\hat{a}^\dagger$) and $\hat{b}$ ($\hat{b}^\dagger$) with a common frequency $\omega_{C}$. These two degenerate modes have an intrinsic loss rate $\kappa$ and are coupled to each other due the backscattering induced by cavity imperfection with constant strength $J$. The multiple two-level atoms with $|g\rangle$ and $|e\rangle$ being the ground state and excited state (with common atomic transition frequency $\omega_{A}$) in the vicinity of the external surface of the cavity interact with modes $\hat{a}$ and $\hat{b}$ via the resonator evanescent fields. We suppose that all atoms have the same dipoles and spontaneous emission rate $\gamma$.      

According to the above scheme, the Hamiltonian for the system can be written in the form,
\begin{eqnarray}
\hat{H} &=&  \sum_{j} \hbar \omega_{A}\sigma_{j}^{+} \sigma_{j}^{-} + \hbar \omega_{C} (\hat{a}^\dagger \hat{a} + \hat{b}^\dagger \hat{b}) + \hbar J (\hat{a}^\dagger \hat{b} + \hat{a}\hat{b}^\dagger) +  \nonumber \\
&&  \sum_{j} \Big[\hbar(g_{twj}\hat{a}\sigma_{j}^{+} + g_{twj}^{*} \hat{a}^{\dagger} \sigma_{j}^{-}) + \hbar(g_{twj}\hat{b}\sigma_{j}^{+} + g_{twj}^{*} \hat{b}^{\dagger} \sigma_{j}^{-}) \Big] + \nonumber \\ 
&& \sum_{i,j} \hbar \Omega_{i,j} (\sigma_{i}^{+} \sigma_{j}^{-} + \sigma_{j}^{+} \sigma_{i}^{-}),
\label{eq:one}
\end{eqnarray}
\noindent where $\sigma_{j}^{+}=|e_j\rangle \langle g_j|$ and $\sigma_{j}^{-}=|g_j\rangle \langle e_j|$ are the raising and lowering operators of the atom $j$. The coherent interaction between atoms and evanescent traveling-wave fields are described by $g_{twj}$ = $g^{twj}_{0} f(\rho, z)\emph{e}^{\pm \textit{i}kx}$, where $\rho$ is the atom-toroid radial distance, $x$ is the atom's position around the circumference of the toroid, $k$ is the vacuum wave vector and $z$ is the vertical coordinate. The $\pm$ refers to the CW ($"+"$) or CCW ($"-"$) propagating modes. $\Omega_{i,j} = \frac{3}{4} (\gamma c^{3}/\omega_{A}^{3} d^{3})(1-3\cos^{2} \theta)$ is the DDI between the $i$th atoms and the $j$th atoms. $d = |\textbf{r}_{i} - \textbf{r}_{j}| \equiv |\textbf{d}|$, where $\textbf{r}_{i}$ is the location coordinate of the $i$th and $\theta$ is the angle between $\textbf{d}$ and the atomic transition dipole moment.

 Following the method developed in Refs. \cite{Dayan,Aoki2}, we can describe this interaction as a function of the normal modes $\hat{A} = (\hat{a}+ \hat{b})/\sqrt{2}$ and $\hat{B} = (\hat{a}-\hat{b})/\sqrt{2}$, and consequently, in the interaction picture,  the Hamiltonian can be rewritten as

 \begin{eqnarray}
\hat{H}_{I} &=&  (\Delta + J)\hat{A}^{\dagger} \hat{A} + (\Delta - J)\hat{B}^{\dagger} \hat{B} + \sum_{i,j} \hbar \Omega_{i,j} (\sigma_{i}^{+} \sigma_{j}^{-} + \sigma_{j}^{+} \sigma_{i}^{-}) + \nonumber \\
&&  \sum_{j} \Big[\hbar g_{Aj}(\hat{A}^\dagger \sigma_{j}^{-} + \hat{A}\sigma_{j}^{+}) - \textit{i} \hbar g_{Bj}(\hat{B}^\dagger \sigma_{j}^{-} -\hat{B}\sigma_{j}^{+}) \Big]
\label{eq:H_I}
\end{eqnarray} 
\noindent where the coupling rates are $g_{Aj}=\sqrt{2} Re(g_{twj}) = g_{0}^{twj}f(\rho , z) \cos (kx)$, $g_{Bj}=\sqrt{2} Im(g_{twj}) = g_{0}^{twj}f(\rho , z) \sin (kx)$ and $\Delta = \omega_{A} - \omega_{C}$. This shows that we can control the coupling strength by adjusting the positions of the atoms. It is  well known that, when the separation between two atoms is much smaller than the optical wavelength, the DDI can be strong. 

From now on, we will consider only the cases of two and three atoms coupled to a microtoroidal cavity and assume two regimes: (i) $kx = n\pi + \pi/4$, where the atoms are coupled simultaneously to the two normal modes ($\hat{A}$ and $\hat{B}$), i. e., symmetric atom-cavity coupling regime, and (ii) the position of $i$th atom at $kx=\pi/4$ and $j$th atom  at $kx=5\pi/4$, so that we can define the asymmetric atom-cavity coupling regime. In the latter case, the distance between the atoms is larger than the resonance wavelength, therefore, DDI can be neglected. Firstly, we will investigate the entanglement dynamics between two atoms with DDI coupled to a microtoroidal cavity. In such a situation, the atomic density operator for two atoms, $\rho_{a}(t)$, that is obtained  by tracing over the resonator field variables, can be written, in the basis $\lbrace |ee\rangle, |eg\rangle, |ge\rangle, |gg\rangle \rbrace$,  as
\begin{equation}
\rho_{a}(t)=\left( \begin{array}{cccc}
A & 0 & 0 & 0 \\ 
0 & B & E^{*} & 0 \\
0 & E & C & 0 \\
0 & 0 & 0 & D \\ 
\end{array} \right)
\label{eq:eq6}
\end{equation}
\noindent where the elements of the density matrix (\ref{eq:eq6}) depend on the initial state. As measure of entanglement between the two atoms we use the negativity for bipartite systems as proposed by Peres and Horodecki \cite{Peres,Horodecki}, that is defined in terms of the negative eigenvalues $\mu^{-}_{i}$ of the partial transposed of the reduced density matrix,
\begin{equation}
\mathcal{N} = -2 \sum_{i} \mu_{i}^{-}.
\label{eq:eq.7}
\end{equation}

The partial transposition  of Eq.(\ref{eq:eq6}), has three eigenvalues. Among them, $\mu^{-}=\frac{1}{2}(D + A - \sqrt{(D-A)^2 + 4E^2})$ that becomes negative under the condition $E^2 > AD$. Using this expression for $\mu^{-}$,  $\mathcal{N}$ can be rewritten as
\begin{equation}
\mathcal{N} = \sqrt{(D-A)^2 + 4E^2} - D - A.
\label{eq:eq11}
\end{equation}  

As proposed by \cite{Peres,Horodecki}, $\mathcal{N}=0$ indicates that the system is separable, $0<\mathcal{N}<1$ means that there is some amount of entanglement and for $\mathcal{N}=1$ the system is maximally entangled.

\section{Effects of initial states on the entanglement dynamics of two atoms}
\label{sec:result1}

In this section, in order to study the relevance of  initial states on the entanglement dynamics between two atoms, we assume that the atoms are initially prepared in a superposition state, $ \cos \theta |ge\rangle + \sin \theta |eg\rangle$. Them, we investigate the atomic entanglement in the process of the time evolution for $\theta = \pi/4$ (initial entangled state) and $\theta = 0$ (initial product state). In the following, we consider two special cases: the symmetric and asymmetric coupling regime.

\subsection{Symmetric coupling regime}

We consider the first case in this subsection, i.e., two two-level atoms with DDI coupled symmetrically to the two WGMs, i.e., $g_{Aj}=g_{Bj} = g$, taking into account the scattering strength between the modes. From the experimental point of view, the DDI can be obtained and controlled by trapping and cooling techniques, reducing the relative distance between the atoms or increasing the intensity of dipole moment for each atom \cite{Harris}. Initially, for an analytical treatment, we consider that the cavity fields are prepared in the vacuum state and the two atoms are prepared in a superposition state $ \cos \theta |ge\rangle + \sin \theta |eg\rangle$. Thus, the evolution of the system state can be described by

\begin{eqnarray}
|\psi (t)\rangle &=& c_1 (t)|eg00\rangle + c_2 (t)|ge00\rangle + c_3(t)(|gg10\rangle + |gg01\rangle)
\end{eqnarray} 
where
\begin{eqnarray}
c_1 (t) = \frac{1}{2}  \left\lbrace {\alpha\emph{e}}^{\textit{i}(J + g\Omega)t/2} cosh(\sqrt{\Delta}t/2)+ \emph{e}^{-\textit{i}g\Omega t} \left[cos(\theta) - sin(\theta) - \frac{\alpha\textit{i} \emph{e}^{\textit{i}(J + 3g\Omega)t/2} (J - g\Omega) sinh(\sqrt{\Delta}t/2)}{\sqrt{\Delta}}   \right] \right\rbrace  
\end{eqnarray}

\begin{eqnarray}
c_2 (t) = \frac{1}{2}  \left\lbrace {\alpha\emph{e}}^{\textit{i}(J + g\Omega)t/2} cosh(\sqrt{\Delta}t/2)+ \emph{e}^{-\textit{i}g\Omega t} \left[-cos(\theta) + sin(\theta) - \frac{\alpha \textit{i} \emph{e}^{\textit{i}(J + 3g\Omega)t/2} (J - g\Omega) sinh(\sqrt{\Delta}t/2)}{\sqrt{\Delta}}   \right] \right\rbrace 
\end{eqnarray}

\begin{eqnarray}
c_3 (t) = \frac{2\alpha g\textit{i} {\emph{e}}^{\textit{i}(J + g\Omega)t/2}  \sinh(\sqrt{\Delta}t/2)}{\sqrt{\Delta}} 
\end{eqnarray}
\noindent with $\Delta = 2gJ\Omega - J^2 - g^2(16 + \Omega^2)$ and $\alpha = \cos (\theta) + \sin (\theta)$. Thus, the time-dependent reduced atomic density operator is given by 
\begin{equation}
\rho_{a}(t)=\left( \begin{array}{cccc}
0 & 0 & 0 & 0 \\ 
0 & |c_1 |^2 & c_1 c_2^{*}   & 0 \\
0 & c_1^{*} c_2 & |c_2|^2  & 0 \\
0 & 0 & 0 & 2|c_3|^2 \\ 
\end{array} \right)
\label{eq:eqxx}
\end{equation}

\noindent In particular, for $\theta=\pi /4 $, e. i., when the two atoms are prepared in a maximally entangled state, the analytical solution for the negativity can be written as 
\begin{multline}
\mathcal{N} = \frac{2}{\Delta} \left\lbrace \frac{1}{4}\left[\Delta \cosh^2 (\sqrt{\Delta}t/2) - (J - g\Omega)\sinh ^2 (\sqrt{\Delta}t/2)\right]^2 +  4 \left[4g^2 \sinh ^2 (\sqrt{\Delta}t/2)\right]^2 \right\rbrace ^{1/2} \\
-\frac{4}{\Delta} \left[4g^2 \sinh^2(\sqrt{\Delta}t/2) \right] 
\label{eq:eq17}
\end{multline}

\noindent From the Eq.(\ref{eq:eq17}) one can see that the negativity is a function of $gt$, $J/g$ and $\Omega/g$. Furthermore, adjusting both the DDI as well as the scattering strength, the atomic system can gain more entanglement. This is because the two atoms indirectly interact with each other via the two WGMs, and directly via the DDI.  In addition, for appropriate values of $\Omega$ and $J$, the negativity can oscillate periodically in function of $gt$ with the period $\frac{4\pi}{\sqrt{\Delta}}g$. Thus, when $\Omega ,J \gg g$, i.e., both the DDI and the scattering strength are much larger than the atom-cavity coupling strength, the negativity tends to an asymptotic value equals 1, which means that the two atoms are trapped in a maximally entangled steady state. These results can be understood in the following way. Under the condition of the above cases, i. e., strong DDI and strong scattering strength, the atoms do not exchange energy with the cavity two-modes, which means that the initial state of the atoms is an eigenstate of the Hamiltonian (Eq.\ref{eq:H_I}). 

\begin{figure}[htp]
\centering
\subfigure[]
{\resizebox{0.35\hsize}{!}{\includegraphics*{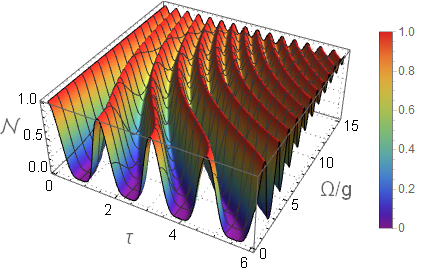}}}\qquad
\subfigure[]
{\resizebox{0.35\hsize}{!}{\includegraphics*{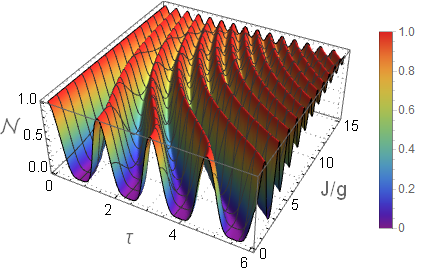}}}\qquad
\subfigure[]
{\resizebox{0.35\hsize}{!}{\includegraphics*{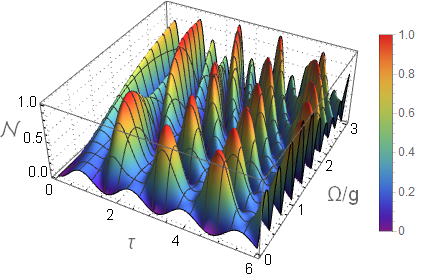}}}\qquad
\subfigure[]
{\resizebox{0.35\hsize}{!}{\includegraphics*{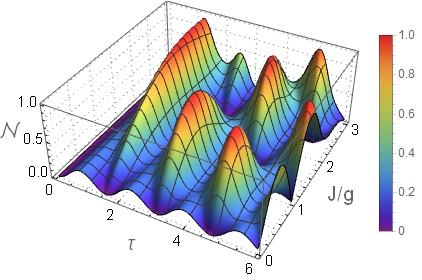}}}
\caption{Negativity as a function of the normalized time $\tau$ and  dipole strength $\Omega$ for $J=0$ in (a)-(c); and scattering strength $J$ for $\Omega=0$ in (b)-(d). In (a) and (b) the atoms are initially prepared in a state $(|ge\rangle + |eg\rangle)/ \sqrt{2}$. In (c) and (d) the atoms are initially prepared in a state $|eg\rangle$. The cavity modes are prepared in a vacuum state.}
\label{fig:fig02}
\end{figure}

In order to show how the initial states, entangled and non-entangled, influences on the entanglement dynamics, the negativity of the atoms is shown in Fig \ref{fig:fig02}. From Fig. \ref{fig:fig02} (a), we can see that the maximam value of negativity of the atoms during the time evolution can be frozen by increasing DDI when the atoms are initially prepared in an  entangled state, as discussed before. From Fig. \ref{fig:fig02}(b), again, one can find that the scattering strength will also contribute to the entanglement of the atoms reaching a maximally entangled stationary state. These results also show that, when the value of $\Omega /g$ or $J/g$ decrease, the negativity also decreases. Specially, for $\Omega/g \rightarrow 0$ or $J/g \rightarrow 0$, the sudden death and birth of entanglement emerge due to the independent interaction of each atom with the fields, causing loss of the atomic coherence in the fields. In contrast when the atoms are initially prepared in a separable state, even when $\Omega /g$ or $J/g$ decrease, the sudden death and birth of entanglement does not occur, as shown in Figs. \ref{fig:fig02}(c) and \ref{fig:fig02}(d). Comparing Figs. \ref{fig:fig02}(a) and \ref{fig:fig02}(b), both displays periodic oscillatory behaviors and even degree of entanglement, independent from the values of $\Omega$ and $J$, so that the atomic system will reach a maximally entangled steady state when the system is prepared in an initial entangled state. However, a completely different result appears when the atoms are initially prepared in a product state, as shown in Figs. \ref{fig:fig02}(c) and \ref{fig:fig02}(d). As expected, it can be seen that DDI contributes in a more significant way on the atomic entanglement than the scattering strength. 
\begin{figure}[htp]
\centering
\subfigure[]
{\resizebox{0.3\hsize}{!}{\includegraphics*{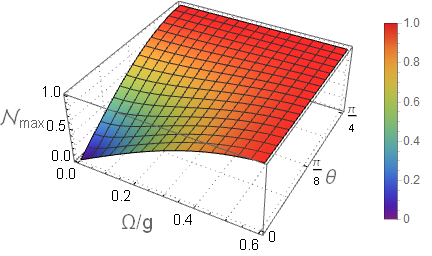}}}\qquad
\subfigure[]
{\resizebox{0.3\hsize}{!}{\includegraphics*{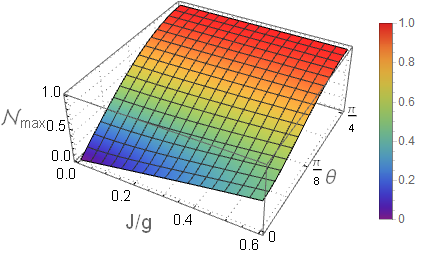}}}
\subfigure[]
{\resizebox{0.3\hsize}{!}{\includegraphics*{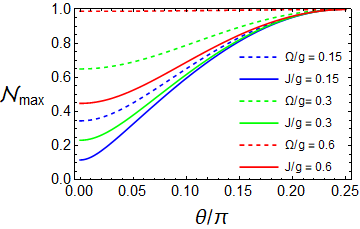}}}
\caption{Maximal atomic negativity as a function of $\theta$ and dipole strength $\Omega$ for $J=0$ in (a); and scattering strength $J$ for $\Omega=0$ in (b). Maximal atomic negativity as a function of $\theta$ for three different values of pairs ($\Omega$,$J$) in (c). In all plots the cavity modes are prepared in a vacuum state.}
\label{fig:fig0444}
\end{figure} 
In order to further explain the above results, let us now show the relationship between the initial states and the system parameters, in the maximal atomic entanglement regime. From Figs. \ref{fig:fig0444}(a) and \ref{fig:fig0444}(b),it can be noted that the maximal negativity of the atoms is more susceptible to variation in scattering strength than the DDI. In Fig. \ref{fig:fig0444}(c), it is  plotted three pairs of cross sections of the surface plots of the Figs. \ref{fig:fig0444}(a) and \ref{fig:fig0444}(b) to make easier a comparison of both cases. One can see that, the influence of $\Omega$ and $J$ on the atomic entanglement depends strongly on the initial state, as can be seen from the two extremes of the entanglement:  $\theta=0$, when the atoms are initially in a separated state and  $\theta=\pi/4$, when the atoms are initially in an entangled state.

\subsection{Asymmetric coupling regime}

In this subsection, we consider $g_{Aj}\neq g_{Bj}$ which indicates that the two atoms are coupled asymmetrically to a WGMs microresonator. In this situation, we have assumed that the position of atom 1 is at $kx = \pi/4$ and atom 2 is at $kx = 5\pi/4$. Thus, the distance between the atoms is much larger than an optical wavelength and the DDI becomes negligible ($\Omega = 0$). We have denoted the coupling constant of the two atoms by the labels $g_{1}$ and $g_2$, respectively. 
 
\begin{figure}[htp]
\centering
\subfigure[]
{\resizebox{0.3\hsize}{!}{\includegraphics*{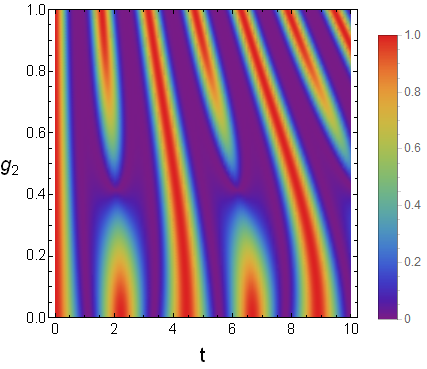}}}\qquad
\subfigure[]
{\resizebox{0.3\hsize}{!}{\includegraphics*{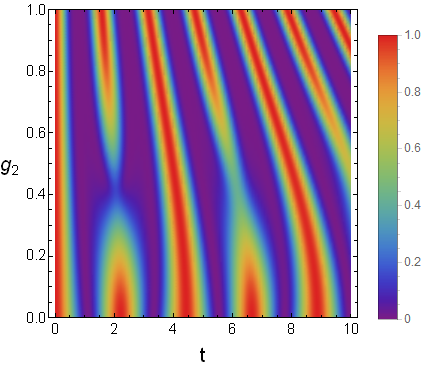}}}\qquad
\subfigure[]
{\resizebox{0.3\hsize}{!}{\includegraphics*{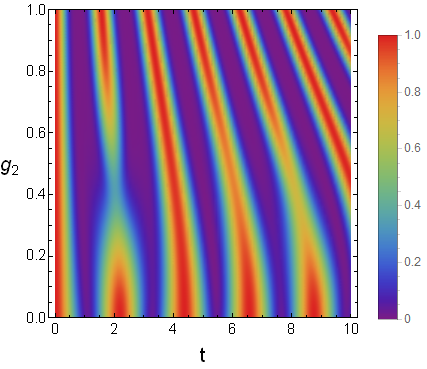}}}\qquad
\subfigure[]
{\resizebox{0.3\hsize}{!}{\includegraphics*{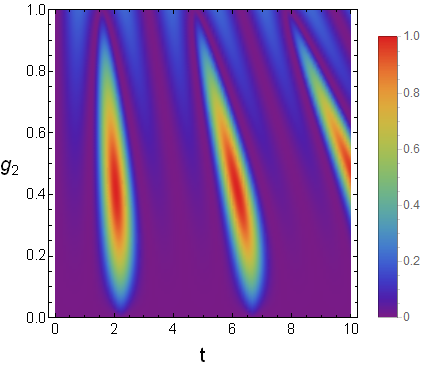}}}\qquad
\subfigure[]
{\resizebox{0.3\hsize}{!}
{\includegraphics*{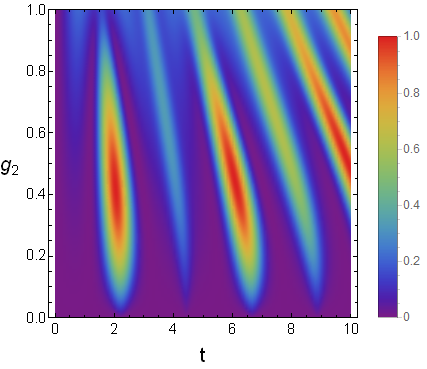}}}\qquad
\subfigure[]
{\resizebox{0.3\hsize}{!}{\includegraphics*{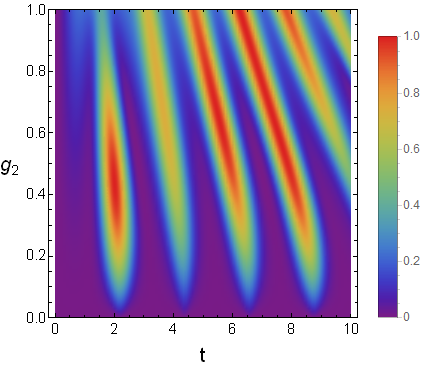}}}
\caption{Negativity as a function of the time $t$ and of the intensity of the constant coupling $g_2$: (a)-(d) $J=0$, (b)-(e) $J=0.2$ and (c)-(f) $J=0.5$, for the cavity modes  prepared in vaccum state. In (a),(b) and (c) the two atoms are initially prepared in an entangled state $(|eg\rangle + |ge\rangle)/\sqrt{2}$ and  in (d),(e) and (f) in a product state $|eg\rangle$. The chosen parameters were $g_{1}=1.0$. }
\label{fig:fig0454}
\end{figure}

In Fig. \ref{fig:fig0454}, we plot the negativity of the atoms versus the time $t$ and intensity of the coupling constant $g_{2}$, for different values of $J$ and $g_{1}=1.0$, when the atoms are initially prepared in an entangled state ($(|eg\rangle + |ge\rangle)/\sqrt{2}$) and a product state ($|eg\rangle$). We can see that, as each atom interacts independently with the two WGMs, the atomic decoherence is being monitored by the detuning between the couplings $g_{1}$ and $g_{2}$. This means that during the time evolution, the atomic entanglement is dependent on the atomic location (via $g_{i}$ values) and scattering strength (via  $J$). It is interesting to note that both in the entangled state and the separate state, the atomic entanglement presents a qualitatively similar behavior, and in this case the coupling constant $g_{1}$ and $g_{2}$ are not able to differentiate the initial state preparation (compare Figs. \ref{fig:fig0454}(c) and \ref{fig:fig0454}(f)). 

\section{Atomic entanglement dynamics with dissipation}
\label{sec:result2}

In this section, in contrast to previous treatments, now we are going to investigate the entanglement dynamics between two two-level atoms coupled to a WGMs microresonators, taking into account  both the cavity decay and atomic spontaneous emission. For the sake of simplicity, we assume the symmetric coupling regime. In this way, considering that both cavity modes have the same decay rate $\kappa$ and both atoms have the same spontaneous emission rate $\gamma$, the system can be described by the master equation

\begin{eqnarray}
\frac{d}{dt}\hat{\rho} &=& -\frac{\textit{i}}{\hbar} [\hat{H}_{I},\hat{\rho}] + \kappa \sum_{O=A,B}[2\hat{O}\hat{\rho}\hat{O}^\dagger - \hat{O}^\dagger \hat{O} \hat{\rho} - \hat{\rho}\hat{O}^\dagger \hat{O}]+ 
\frac{\gamma}{2} \sum_{i=1,2}[2\sigma_{i}^{-}\hat{\rho}\sigma_{i}^{+} - \sigma_{i}^{+}\sigma_{i}^{-} \hat{\rho} - \hat{\rho}\sigma_{i}^{+}\sigma_{i}^{-}].
\end{eqnarray}

\begin{figure}[htp]
\centering 
\subfigure[]
{\resizebox{0.4\hsize}{!}{\includegraphics*{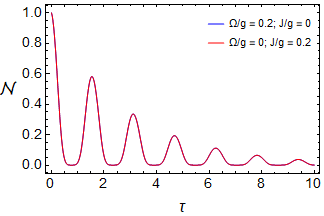}}}\qquad
\subfigure[]
{\resizebox{0.4\hsize}{!}{\includegraphics*{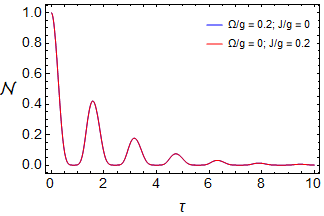}}}\qquad
\subfigure[]
{\resizebox{0.4\hsize}{!}{\includegraphics*{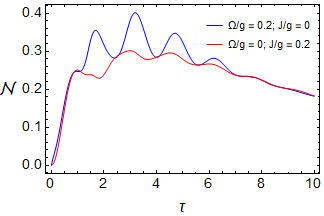}}}\qquad
\subfigure[]
{\resizebox{0.4\hsize}{!}{\includegraphics*{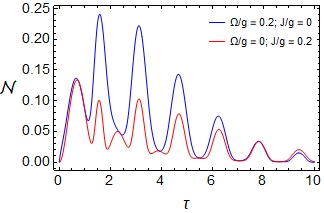}}}\qquad
\caption{Negativity as a function of the normalized time $\tau$ when the atoms are prepared in an  entangled state in (a)-(b), and in a separate state in (c)-(d). The chosen parameters were: $\Omega/g=0.2, J/g=0$ (blue line) and $\Omega/g=0, J/g=0.2$ (red line) and for (a)-(c) $\kappa=0.5g$ and $\gamma=0.1g$, and for (b)-(d) $\kappa=0.1g$ and $\gamma=0.5g$.}
\label{fig:fig099}
\end{figure}

 It is evident that the analytical solution of the master equation is quite difficult to achieve. In order to explicitly elucidate the initial state influences on the entanglement dynamics, we discuss this case via numerical simulations. 
In Fig.\ref{fig:fig099}, we plot the negativity of the atoms as a function of the normalized time $\tau$ for $\Omega/g=0.2, J/g=0$ (blue line) and $\Omega/g=0, J/g=0.2$ (red line), (a)-(c) $\kappa=0.1g$ and $\gamma=0.5g$, (b)-(d) $\kappa=0.5g$ and $\gamma=0.1g$. 
The atomic entanglement, as it is known, decreases with the intrinsic loss rate of the cavity and of the atomic spontaneous emission (see Fig. \ref{fig:fig099}).
From Figs.\ref{fig:fig099}(a) and \ref{fig:fig099}(b), we can see that, for different values of $\Omega$ and $J$, the entanglement decay has no differences, because the atoms are initially in an entangled state, as discussed previously. However, when atoms are in an initial product state, the entanglement dynamics is dependent on the values of $\Omega$ and $J$, as shown in Figs.\ref{fig:fig099}(c) and \ref{fig:fig099}(d). Moreover, from the behavior shown in the figures, it is evident that the entanglement loss due to the dissipative factors can be suppressed at certain time intervals by appropriately adjusting the values of $\Omega$ and $J$ \cite{Jin}.
     
The difference between Figs.\ref{fig:fig099} (a) and \ref{fig:fig099} (b) is that the atomic entanglement decreases quickly when the atomic decay rate is greater than the cavity loss.
Thus, we can infer that the cavity loss rate is not a dominant factor for the degradation of the atomic entanglement. As the cavity modes and the atoms are decoupled one from another the atomic entanglement dynamics are favorable. Similar characteristics are obtained when the atoms are prepared in a separate state. (See Figs.\ref{fig:fig099}(c) and \ref{fig:fig099}(d)).

\section{Dynamics of tripartite atomic entanglement with dissipation}
\label{sec:multipartite}

\begin{figure}[htp]
\centering
\subfigure[]
{\resizebox{0.4\hsize}{!}{\includegraphics*{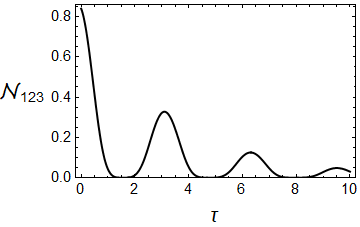}}}\qquad
\subfigure[]
{\resizebox{0.4\hsize}{!}{\includegraphics*{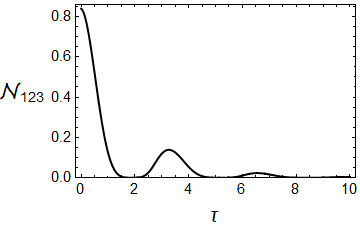}}}
\subfigure[]
{\resizebox{0.4\hsize}{!}{\includegraphics*{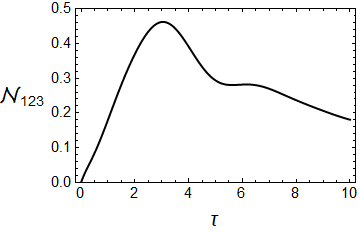}}}\qquad
\subfigure[]
{\resizebox{0.4\hsize}{!}{\includegraphics*{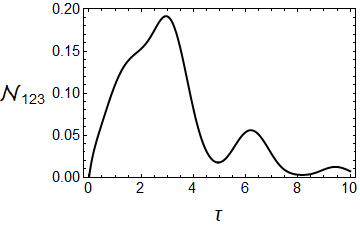}}}
\caption{Tripartite negativity as a function normalized time $\tau$. The atoms are prepared in a state $(|egg\rangle + |geg\rangle + |gge\rangle)/\sqrt{3}$ in (a)-(b) and product state $|egg\rangle$ in (c)-(d). The cavity modes are in vacuum state. The parameters are chosen as (a)-(c)$\gamma = 0.1g$ and $\kappa=0.5g$; (b)-(d) $\gamma = 0.5g$ and $\kappa=0.1g$ for $\Omega = J =0.1g$.}
\label{fig:fig17}
\end{figure}

The bipartite entanglement described in the previous sections leads us toward the main point of our discussion, the multipartite system, that in our case is the tripartite system. As is well known, the tripartite system entanglement plays an important role in implementing distributed quantum communication and quantum computation \cite{Amico}. Also, in the tripartite case, the generation of entanglement between three atoms coupled to the two WGM microresonator is quite sensitive to the preparation of the initial states. Thus, to describe the dynamics of the tripartite entanglement, we use the tripartite negativity as defined by \cite{Sabin}
\begin{equation}
\mathcal{N}_{123} = (\mathcal{N}_{1-23} \mathcal{N}_{2-13} \mathcal{N}_{3-12})^{1/3}
\end{equation}
where the bipartite negativities ($\mathcal{N}_{1-23}$, $\mathcal{N}_{2-13}$ and $\mathcal{N}_{3-12}$) are defined as in Eq. \ref{eq:eq.7}. For the sake of simplicity, we assume that the three atoms are coupled symmetrically to the two WGMs, e.g., $g_{1}=g_{2}=g_{3}=g$ and with the same DDI ($\Omega_{1,2} = \Omega_{2,3} = \Omega_{1,3} = \Omega$). In Fig. \ref{fig:fig17}, we plot the tripartite negativity as a function of the normalized time $\tau$ for the initial state of type $W$, i.e., $(|egg\rangle + |geg\rangle + |gge\rangle)/\sqrt{3}$ in (a)-(b), and for initial product state $|egg\rangle$ in (c)-(d), when $\Omega=J=0.1g$. The chosen parameters were: (a)-(c) $\gamma =0.1g$ and $\kappa=0.5g$; (b)-(d)$\gamma =0.5g$ and $\kappa=0.1g$. From Figs. \ref{fig:fig17}(a) and \ref{fig:fig17}(b), analogous to bipartite entanglement, the cavity leaking  when compared to the atomic decay rate, is not a dominant factor in the entanglement loss, even when the atoms are in an initial product state (see Figs.\ref{fig:fig17}(c) and \ref{fig:fig17}(d)).  

Furthermore, even that the degree of tripartite entanglement being lower rate compared to bipartite entanglement (see Fig.\ref{fig:fig099}), one can see that, the time evolution of the tripartite negativity have an enhancement in certain time intervals (see Figs.\ref{fig:fig17}(c) and \ref{fig:fig17}(d)). 
However, the difference between the degree of tripartite entanglement and bipartite entanglement is already well known and is consistent with the so-called entanglement monogamy phenomenon \cite{Coffman}.
Basically, the monogamic entanglement is characterized by the quantitative measure of entanglement described by bipartite quantum entanglement of the quantum system.
In fact, the more information about the system that can be obtained through a measure of entanglement, the greater the efficiency to describe the degree of the system entanglement and the degree of information on it.
In general, for quantum information processing it is necessary to work with more than a few nodes. This means that the loss of quantum capability is unavoidable. But, although the increases of the number of systems promote a decrease of entanglement as shown in the Fig.\ref{fig:fig17}, as results of monogamy effects, in the system here studied, the decrease is not so much pronounced that allow us to bet in this kind of system for the transference of quantum information.

\section{Conclusions}
\label{sec:conclusion}

In this paper, we have explored how the preparation of the initial state influences on the entanglement dynamics between two and three two-level atoms with DDI coupled symmetrically or asymmetrically to a high-$Q$ WGMs microtoroidal cavity. We have considered two types of initial states, entangled and product state between the atoms. We found that the effects of the initial state on the atomic entanglement are remarkable. Significantly, for the initial entangled states and a symmetrical regime, by adjusting both the scattering strength and DDI, the sudden death and birth, as well as the freezing entanglement, can be obtained. Apart from that, we have studied the influence of different initial states and other system parameters on the atomic entanglement. Our results show that the maximal entanglement is more susceptible to variation in scattering strength than the DDI.  
For the asymmetric coupling regime, it is clearly shown that the maximum entanglement between the two atoms depends on the atomic location and can be improved by adjusting the scattering strength.
In addition, entanglement decay obviously occurs due to dissipative factors, however, the effect of the cavity loss in the current scheme is not a dominant factor for the degradation of atomic entanglement. When the cavity modes are decoupled from the atoms, the entanglement dynamics is more favorable for $\kappa \gg \gamma$, even for the case of two or three atoms coupled to a microtoroidal cavity. Similar results are obtained for the case in that nitrogen-vacancy centers are coupled to a microtoroidal cavity. Therefore, our results may contribute to a better understanding of the bipartite and tripartite quantum correlation that is of great interest in quantum information processing.


%
%

\end{document}